\documentclass{ws-procs9x6}

\allowdisplaybreaks[3]

\begin{document}

\title{Hadronic aspects of exotic baryons}

\author{E.~Oset, S. Sarkar, M.J. Vicente Vacas, V. Mateu,
}

\address{Departamento de F\'\i sica Te\'orica and IFIC,
Centro Mixto Universidad de Valencia-CSIC,
Institutos de Investigaci\'on de Paterna, Aptd. 22085, 46071
Valencia, Spain}

\author{T.~Hyodo, A.~Hosaka}

\address{Research Center for Nuclear Physics (RCNP),
Ibaraki, Osaka 567-0047, Japan}  

\author{F. J. Llanes-Estrada}

\address{Departamento de F\'\i sica Te\'orica I, Universidad Complutense, Madrid,
Spain}  

\maketitle

\abstracts{In this talk I look into three different topics, addressing first 
the possibility
that the $\Theta^+$ is a bound state of $K \pi N$, exploiting the results of this
study to find out the contribution of two meson and one baryon components in the
baryon antidecuplet and in the third place I
present results on a new resonant exotic baryonic state which appears as
dynamically generated by the Weinberg Tomozawa $\Delta K$ interaction.}
 
\section{Is the $\Theta ^+$ a $K \pi N$ bound state?}
The experiment by LEPS collaboration
at SPring-8/Osaka~\cite{nakano}
has found a clear signal for an $S=+1$ positive charge resonance
around 1540 MeV. The signal is also found in many other experiments and not 
found in some experiments at high energy, and is subject of intense study in
different labs to obtain higher statistics. A list of papers on the issue can be
found in \cite{hyodolist}.
At a time when many  low energy baryonic resonances are well described as being
dynamically generated as meson baryon quasibound states within chiral 
unitary
approaches \cite{Kaiser:1995eg,Oset:1997it,oller,Jido:2003cb,Nieves:2001wt} it 
looks tempting to
investigate the
possibility of this state being a quasibound state of a meson and a baryon 
or two  mesons and a baryon.  Its nature as a $K N$ s-wave state is easily
ruled 
out since the interaction is repulsive. 
$KN$ in a p-wave, which is attractive, is too weak to 
bind. The next logical possibility is to 
consider a quasibound state of $K 
\pi N$, which in s-wave would naturally correspond to spin-parity 
$1/2^+$, the quantum numbers suggested in
\cite{Diakonov:1997mm}.
Such an idea has already been put forward in \cite{Bicudo:2003rw} where a  
study of the interaction of the three body system is conducted in the 
context of chiral quark models. A
more detailed work is done in \cite{felipe}, which we summarize here.

Upon considering the possible structure of $\Theta^+$ 
we are guided by the  experimental observation
\cite{Barth:2003ja} that the
state is not produced in the $K^+ p$ final state. This would
rule out the possibility of the $\Theta$ state having 
isospin I=1. Then we accept the $\Theta^+$ to be an I=0 state. 
As we couple a pion and a kaon to the nucleon to form such state,
a consequence is that the $K \pi$ substate must combine to I=1/2 and 
not I=3/2.
 This is also welcome dynamically since the s-wave $K \pi$ interaction in
I=1/2 is attractive (in I=3/2 repulsive) \cite{Oller:1998hw0}.
The attractive interaction in I=1/2 is very strong and gives
rise to the dynamical generation of the scalar $\kappa$ resonance around 
850 MeV and with a large width \cite{Oller:1998hw0}.

In order to determine the possible $\Theta^+$ state we search for poles of
the $K\pi N \to K\pi N$ scattering matrix. To such point we construct the
series of diagrams of fig. 1,
\begin{figure}[tbp]
    \centering
    \includegraphics[width=13cm,clip]{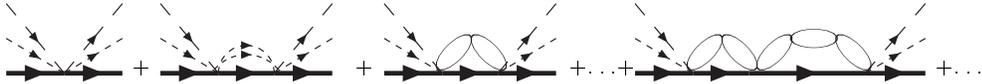}
\caption{\label{trickedamplitude1} Diagrams considered in the $\kappa N$
interaction.} 
\end{figure}
where we account explicitly for the $K\pi$ interaction by constructing
correlated $K\pi$ pairs and letting the intermediate $K\pi$ and nucleon 
propagate. 
This requires a kernel for the two meson-nucleon interaction which we
now address. 
We formulate the meson-baryon lagrangian in terms of
the SU(3) matrices, $B$, $\Gamma_\mu$, $u_{\mu}$
 and the implicit meson matrix 
$\Phi$ standard in ChPT \cite{Bernard:1995dp},
\begin{equation}  
{\it L}= {\rm Tr} \left( \overline{B} i\gamma^\mu \nabla_\mu B \right)
- M_B {\rm Tr}\left(\overline{B}B  \right) \\
+ \frac{1}{2} D {\rm Tr} \left( \overline{B} \gamma^\mu \gamma_5 \left\{
u_\mu,B\right\}   \right) +
\frac{1}{2} F {\rm Tr} \left( \overline{B} \gamma^\mu \gamma_5 \left[
u_\mu, B \right] \right)
\end{equation}
with the definitions in \cite{Bernard:1995dp}.

First there is a contact three body force simultaneously involving 
the pion, kaon and nucleon, which can be derived from the meson-
baryon Lagrangian  term containing the covariant derivative $\nabla_{\mu}$.

By taking the isospin I=1/2 $\kappa$ states
and combining them with the nucleon we generate 
I=0,1 states
which diagonalize the scattering matrix associated to $t_{mB}$ and we find
 that the interaction in the I=0 channel is
attractive, while in the I=1 channel is repulsive. 
This would give chances to the $\kappa N$ $t$-matrix to develop a pole in 
the bound region, but rules out the I=1 state. 

The series  of terms of Fig. 1 might lead to a bound state of
$\kappa N$ which would not decay since the only intermediate channel
is made out of $K \pi N$ with mass above the available energy. 
The decay into $K N$ observed experimentally can
be taken into account explicitly and this and other diagrams accounting 
for the interaction of
the mesons with the other meson or the nucleon are taken into account in the
calculations \cite{felipe}.

What we find at the end is that, in spite of the attraction found, this
interaction is not enough to bind the system, since we do not find a pole below
the $K 	\pi N$ threshold.
 In order to quantify this second statement 
we increase artificially the potential $t_{mB}$ by adding to it a quantity
which leads to a pole around $\sqrt{s}=1540 \ MeV$ with a width of
around $\Gamma=40 \ MeV$. This is accomplished by 
adding an attractive potential around five or six times bigger than
the existing one.   
 We should however note that we have not exhausted all possible sources of
three body interaction since only those tied to the Weinberg Tomozawa term have
been considered. We think that some  more work in this direction should be 
still encouraged and there are already some steps given in
\cite{Kishimoto:2003xy}.

\section{Coupling of the $\Theta^+$ to $K \pi N$}
  Although not enough to bind the $K \pi N$ system, the interaction has proved
 attractive in L=0 and I=0. This, together with the proximity of the $\Theta^+$
 mass to the  $K \pi N$ threshold ( 30 MeV) suggests that the $\Theta^+$ should
 have a non negligible  $K \pi N$ component in its wave function.  The 
 procedure followed in \cite{colajap} to find out the contribution to
 the binding is the following:
 1) one assumes that the $\Theta^+$ belongs to the standard antidecuplet of
 baryons suggested in \cite{Diakonov:1997mm}. 2) The $N^*(1710)$ is assumed to have a
 large component corresponding to this antidecuplet. 3) From the large decay of
 the $N^*(1710)$ into $\pi \pi N$, both in s-wave and p-wave, we extract the
 strength for two SU(3) invariant phenomenological potentials which allow us to
 extend the coupling to different meson meson baryon components of all baryons
of the antidecuplet.  4) A selfenergy diagram is constructed with two vertices
from these Lagrangians and two meson and a baryon intermediate states. 5)
Regularization of the loops is done with a cut off similar to the one needed in
the study of the $\bar{K} N$ interaction \cite{Oset:1997it} and this leads to
attractive selfenergies of the order of 100-150 MeV. At the same time one finds
an energy splitting between the different members of the antidecuplet of the
order of 20 MeV, or 20 percent of the empirical values, with the right ordering
demanded by the Gell-Mann-Okubo rule, and hence a maximum binding for the  
$\Theta^+$. 

  This finding means that detailed studies of the $\Theta^+$ should take into
consideration this important component of $K \pi N$ which helps produce extra
binding for the $\Theta^+$, one of the problems faced by ordinary quark models.

  The finding of this work has repercussions in the selfenergy of the 
$\Theta^+$ in nuclei.  Indeed, as found in \cite{Cabrera:2004yg}, when one takes
into account the pionic medium polarization, exciting $ph$ and $\Delta h$
components with the pion, the mechanism leads to an extra attraction in the
medium which is of the order of 50-100 MeV at normal nuclear matter density.
This, together with the other finding of a very small imaginary part of the
selfenergy, leads to levels of the $\Theta^+$ which are separated by energies
far larger than the width of the states.  This makes it a clear case for
experimental observation and suggestions of reactions have already been made
\cite{Nagahiro:2004wu}.

\section{A resonant $\Delta K$ state as a dynamically generated exotic baryon}
Given the success of the chiral unitary approach in generating dynamically low
energy resonances from the interaction of the octets of stable baryons
and the pseudoscalar mesons,   in 
\cite{Kolomeitsev:2003kt} the
interaction of the decuplet of $3/2^+$ with the octet of pseudoscalar mesons 
was studied and shown to
lead to many states which were associated to experimentally well 
established resonances. The purpose of the present work is to show that this
interaction leads also to a new state of positive strangeness, with $I=1$ 
and $J^P =3/2^-$, hence, an exotic baryon described in terms of a resonant
state of a $\Delta$ and a $K$.
  
    The lowest order chiral Lagrangian for the interaction of the baryon 
decuplet with the octet of pseudoscalar mesons is given by \cite{Jenkins:1991es}
\begin{equation}
L=i\bar T^\mu D_{\nu} \gamma^{\nu} T_\mu -m_T\bar T^\mu T_\mu
\label{lag1} 
\end{equation}
where $T^\mu_{abc}$ is the spin decuplet field and $D^{\nu}$ the covariant derivative
given  in \cite{Jenkins:1991es}. The identification of the physical decuplet
states with the $T^\mu_{abc}$ can be seen in \cite{sarkar}.

For strangeness $S=1$ and charge $Q=3$ there is only one channel $\Delta^{++}
K^+$ which has $I=2$. For $S=1$ and $Q=2$ there are two channels 
$ \Delta^{++}K^0$ and $\Delta^{+}K^+$. From these one can extract the transition
amplitudes for the $I=2$ and $I=1$
combinations and we find \cite{sarkar}
\begin{equation}
V(S=1,I=2)=\frac{3}{4f^2}(k^0+k^{\prime 0}); ~~~~~V(S=1,I=1)=-\frac{1}{4f^2}
(k^0+k^{\prime 0}),
\label{pot}
\end{equation}
where $k(k^{\prime})$ indicate the incoming (outgoing) meson momenta. 
These results indicate that the
interaction in the $I=2$ channel is repulsive while it is attractive in $I=1$.

The use of $V$ as the kernel of the Bethe Salpeter equation
\cite{Oset:1997it}, or the N/D unitary approach of \cite{oller} both lead to the
scattering amplitude 
\begin{equation}
t=(1-VG)^{-1}V
\label{LS}
\end{equation}
 In eq. (\ref{LS}), $V$ factorizes on shell
\cite{Oset:1997it,oller} and $G$ stands for the loop function of the meson and baryon
propagators, the expressions for those being given in \cite{Oset:1997it} for a cut off
regularization and in \cite{oller} for dimensional regularization. 
  
  Next we fix the scale of regularization by determining the cut off, $q_{max}$,
in the loop function of the meson and baryon propagators in order to reproduce 
the resonances for other 
strangeness and isospin channels. They are one resonance in
$(I,S)=(0,-3)$, another one in $(I,S)=(1/2,-2)$ and another one in
$(I,S)=(1,-1)$. The last two appear in \cite{Kolomeitsev:2003kt} around 1800 MeV and 1600 MeV and they are
identified with the $\Xi(1820)$ and $\Sigma(1670)$.  We obtain the same results
as in \cite{Kolomeitsev:2003kt} using a cut off $q_{max}=700$ MeV.

  With this cut off we explore the analytical properties of the amplitude for
$S=1$, $I=1$ in the first and second Riemann sheets. First we see that there is
no pole in the first Riemann sheet.  

  Next we explore the second Riemann sheet which we obtain by changing the sign
  of the momentum in the expresssion for the meson baryon loop function.

 We find a pole at $\sqrt{s}=1635$ MeV in the second Riemann
sheet. 
\begin{figure}[tbp]
    \centering
    \includegraphics[width=10cm,clip]{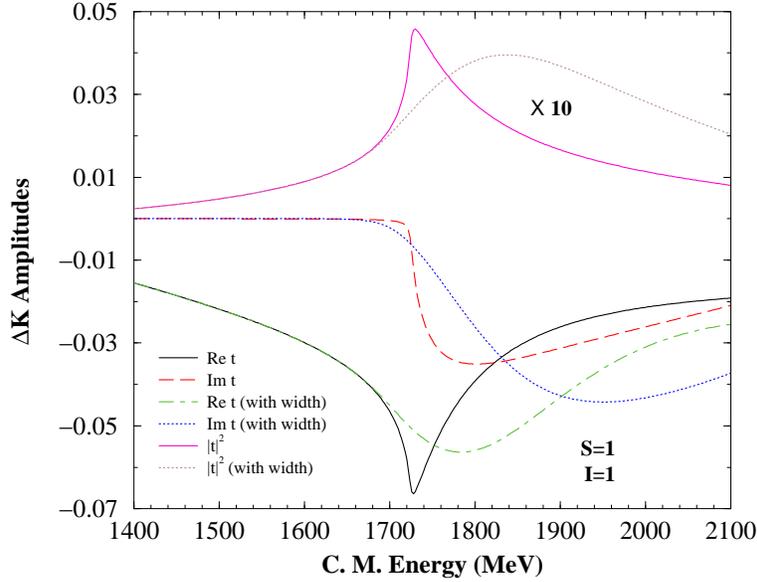}   
\caption{Amplitudes for $\Delta K\rightarrow\Delta K$ for $I=1$}
\end{figure}

The situation in the scattering matrix is revealed in fig. 2  which shows
the real and imaginary part of the $K \Delta$ 
amplitudes for the case of $I=1$.  For the case of
$I=2$ the imaginary part follows the ordinary behaviour of the opening of a 
threshold,
growing smoothly from threshold. The real part is also smooth.
For the case of $I=1$, instead, the strength of the imaginary part is stuck to
threshold as a reminder of the existing pole in the complex plane, growing very
fast with energy close to threshold.  The real part
has also a pronounced cusp at threshold, which is also tied to the same
singularity.

 In figure 2 and we see that the peak around
threshold becomes smoother and some strength is moved to higher energies when we
consider the width
of the $\Delta$ in the intermediate states.  Even
then, the strength of the real and imaginary parts in the $I=1$ are much larger
than for $I=2$. The modulus squared of the amplitudes shows some
peak behavior around 1800 MeV  in the case of $I=1$, while it is small and has no
structure in the case of  $I=2$.

We propose the study 
of the
following reactions:  1) $pp \to \Lambda \Delta^+ K^+$, 2) $pp \to \Sigma^- \Delta^{++}
K^+$, 3) $pp \to \Sigma^0 \Delta^{++}K^0$. In the first case the $\Delta^+ K^+$ state
produced has necessarily $I=1$.  In the second case the $\Delta^{++}K^+$ state has
$I=2$. In the third case the $\Delta^{++}K^0$ state has mostly an $I=1$ component.
The
experimental confirmation of the results found here through the study of the
$\Delta K$ invariant mass distribution in these reactions would give 
support to this new exotic baryonic state which stands as a
resonant $\Delta K$ state. 

\section{Acknowledgments}
This work is partly supported by DGICYT contract number BFM2003-00856,
 the E.U. EURIDICE network contract no. HPRN-CT-2002-00311 and the Research
 Cooperation program of the  JSPS and the  CSIC.

\end{document}